Superconducting dome due to the Fano-Feshbach shape resonance in artificial high-$T_C$ superlattices

Gaetano Campi[1,2,*], Andrea Alimenti[3,4,†], Danielle Villa[5,‖], G. Alexander Smith[5, §], Fedor F. Balakirev[5, 6¶], Luis Balicas[7, #], Andrea Perali[8 **], Sergio Caprara[9 §], Gennady Logvenov[10 ‡], Antonio Bianconi[1,2, ‖‖]

[1]Institute of Crystallography, National Research Council, CNR, Via Salaria Km 29.3, 00015 Monterotondo Rome, Italy
[2]Rome International Center for Materials Science Superstripes RICMASS, Via dei Sabelli 119A, 00185 Rome, Italy
[3]Department of Industrial, Electronic and Mechanical Engineering, Roma Tre University, Via Vito Volterra 62, 00146 Rome, Italy
[4]Istituto Nazionale di Fisica Nucleare INFN, Sezione Roma Tre, Via della Vasca Navale 84, 00146 Rome, Italy
[5]National High Magnetic Field Laboratory (NHMFL), Los Alamos National Laboratory (LANL), Los Alamos, New Mexico 87545, USA,
[6]Laboratoire National des Champs Magnétiques Intenses, 31400 Toulouse France
[7]Department of Physics and Astronomy, Baylor University, Waco, Texas, USA
[8]CQM group, School of Pharmacy, Physics Unit, University of Camerino, 62032 Camerino, Italy
[9] Department of Physics, Sapienza University, Ple. Aldo Moro 5, 00185 Roma, Italy
[10]Max Planck Institute for Solid State Research, Heisenbergstraße 1, 70569 Stuttgart, Germany

Corresponding authors: Gaetano Campi*, Andrea Alimenti†, Antonio Bianconi ‖‖
Authors Email:
* gaetano.campi@cnr.it,
† andrea.alimenti@uniroma3.it,
‖ dvilla@lanl.gov,
§ gasmith@lanl.gov,
¶ fbalakirev@gmail.com,
# Luis_Balicas@baylor.edu,
** andrea.perali@unicam.it,
§ sergio.caprara@uniroma1.it,
‡ g.logvenov@fkf.mpg.de,
‖‖ antonio.bianconi@ricmass.eu

## Abstract

In this work we provide compelling experimental validation of the Bianconi Perali Valletta (BPV) theory predicting a "superconducting dome" based on a quantum material design of Artificial High $T_C$ Superlattices (AHTS) made with a selected nanoscale heterostructure geometry. These AHTS are SNSN superlattices of quantum wells of period *d*, composed of first units, superconducting doped Mott insulator layers with Rashba spin orbit coupling (S) of thickness L, intercalated by second units, normal metal spacers (N). In these superlattices, grown by molecular beam epitaxy (MBE), the experimental "superconducting dome" is obtained by material quantum design changing the chemical potential via the quantum geometrical factor L/d which tunes the Fano-Feshbach *shape resonance* in the pair transfer between superconducting gaps in the BCS regime and different gaps in the BEC-BCS crossover. Here we present a systematic magneto-transport study of AHTS artificial superlattices across the full doping range of the superconducting dome, from the deeply underdoped to the overdoped regime, using pulsed magnetic fields up to 72 T. By varying the $L/d$ ratio, we tune the effective hole concentration $\delta$=0.45(1 - $L/d$) and map the evolution of the resistive transitions, the upper critical field $\mu_0 H_{C2}(T)$, and the Ginzburg-Landau coherence length $\xi_0$.

## 1. Introduction

While the critical temperature of conventional BCS superconductors does not depend on the chemical potential, the signature of unconventional high-temperature superconductors is a phase diagram showing a “superconducting dome” which describes how the critical temperature $T_C$ rises, peaks, and then falls as a material's chemical potential is varied. In the frame of the traditional paradigm of theoretical models assuming a homogeneous lattice and single gap superconductivity it has been assigned to a quantum critical point, on the contrary an alternative theoretical paradigm, called since 1996 Bianconi-Perali-Valletta (BPV) theory, assigns the origin of the “superconducting dome” to a Fano-Feshbach *shape resonance* between first, superconducting gaps in the BCS regime and second superconducting gaps in the BEC-BCS regime in a multigap superconductivity scenario, where the functional nanoscale heterostructure driven by nanoscale arrested phase separation of the lattice structure plays a key role.
Understanding resonant multigap superconductivity in the BEC-BCS crossover emerging in doped Mott insulators remains one of the most enduring challenges in condensed matter physics. Among cuprate perovskites $La_2CuO_4$ (LCO) is considered the archetypal antiferromagnetic Mott insulator, where the superconductivity dome appears in *doped Mott insulator* $La_{2-x}Sr_xCuO_4$ through chemical substitution around the insulator-to-metal transition (IMT) in the normal phase at $x \approx 0.18$.
In complex quantum materials arrested nanoscale phase separation of doped strongly correlated systems, can give rise to the coexistence of chemically overdoped metallic regions, dopants-free Mott insulator regions, and competing electronic orders, such as charge-density waves (CDWs), over length scales of only a few nanometers [1–17]. Recently, similar nanoscale texturing has been reported in strongly correlated oxide $NdNiO_3$ [18], iron-based superconductors [19–20], bismuth chalcogenides, and other quantum complex materials [21–26]. A universal tendency was identified for multiband doped Mott insulators toward intrinsic arrested nanoscale phase separation near Lifshitz electronic topological transitions [27-30], stripe phases, and charge-density waves (CDW) competing with high $T_C$ superconductivity as a function of lattice strain [31-36].

A route to high-temperature superconductivity emerging in chemically doped Mott insulators which is fundamentally different from the “trial and error” methods, was recently opened by using quantum materials design principles for the fabrication of nanoscale artificial heterostructures which can be used to test predicted theoretical scenarios. First, we avoid chemical doping, which, beyond changing material functionality, inevitably generates a landscape of impurity scattering, local lattice distortions, and internal chemical pressure which could mask the intrinsic features of the low-temperature insulator-to-metal transition (IMT) in doped Mott insulator and we first focus on the critical doping for the low-temperature IMT transition [37-39]. We use a two-dimensional doping [40-42] of stoichiometric $La_2CuO_4$ Mott insulating layers which are periodically intercalated between metallic non superconducting overdoped layers [43–47]. This approach effectively decouples the introduction of carriers in $La_2CuO_4$ from the generation of chemical disorder, offering a structurally cleaner platform to investigate the intrinsic physics of the superconducting state in a superconducting doped Mott insulator. In these heterostructures, the carrier density in the stoichiometric Mott-insulating blocks is due to quantum-confined interfacial space charge transferred from the adjacent metallic units, generating a superconducting state of the Mott insulator confined in quantum wells.

The theoretical framework guiding the design of such artificial structures is the Bianconi–Perali–Valletta (BPV) theory [48-55] for heterostructures made of nanoscale building blocks [48], where the superconducting dome is due to a Fano-Feshbach resonance in a multigap superconductor at the BEC-BCS crossover. The “rising edge” of the dome is in the BEC-like regime and the “drop edge” of the dome is in low-$T_C$ BCS regime. This predictive quantum design paradigm allows engineering high-$T_C$ superconductivity from the interfacial charge in the presence of Rashba spin–orbit coupling due to internal interface electric field in heterostructures made of nanoscale quantum building blocks from first principles. Unlike empirical trial-and-error approaches, the BPV theory that includes spin-orbit interaction identifies the geometrical conformational parameter $L/d$, defined as the ratio between the thickness $L$ of the superconducting quantum-well layers and the superlattice period $d$=$L$+$W$ (where $W$ is the thickness of the metallic spacer) as the key control variable for the superconducting critical temperature. The theory predicts that the optimal superconductivity occurs at the optimal ratio $L/d$=2/3, where the system undergoes a Lifshitz electronic topological transition that drives a Fano–Feshbach resonance between subbands in the quantum wells. At the Fano-Feshbach resonance, the emergence of multiband, two-gap superconductivity leads to the enhancement of $T_C$ and a long Ginzburg coherence length. Here, we validate this prediction experimentally in LSCO/LCO superlattices grown via molecular beam epitaxy, where the engineered samples confirm the theoretical expectations of a multiband superconductor with two distinct energy gaps.

Despite these advances, a comprehensive investigation of the upper critical field $\mu_0 H_{C2}$ across the entire superconducting dome of artificial high-TC superlattices (AHTS) is lacking. While previous studies have established the multiband character and the Fano–Feshbach resonance in individual compositions near optimal doping, a systematic mapping of $\mu_0 H_{C2}$(T) from the deeply underdoped to the overdoped regime has yet to be reported. The behaviour of $\mu_0 H_{C2}$(T) provides crucial information not only on the superconducting phase boundary, but also on the pairing mechanism, the possible multiband character of the condensate, and the interplay between superconductivity and the normal-state electronic correlations. In conventional chemically doped cuprates, the underdoped side of the dome is strongly affected by a competing CDW order [36-37], which suppresses both $T_C$ and $\mu_0 H_{C2}$, obscuring the intrinsic superconducting properties. Whether the same phenomenology persists in the cleaner environment of artificial heterostructures, or whether new physics emerges, is an open question of fundamental importance.

We compare our results with the archetypal superconducting unconventional high-temperature superconducting cuprate $La_{2-x}Sr_xCuO_4$, which exhibits a superconducting dome by tuning $0.05 < x < 0.3$, with maximum $T_C$ at x ≈ 0.15 and low temperature insulator-to-metal transition (IMT) at x ≈ 0.18. In this work we report the superconducting dome in artificial high-TC superlattices (AHTS) of stoichiometric $La_2CuO_4$ (LCO) quantum wells of thickness $L$, intercalated with overdoped $La_{1.55}Sr_{0.45}CuO_4$ (LSCO) spacers of thickness $W$, with period d=$L$+$W$, controlled geometrically by L/d, without chemical substitution in the superconducting LCO layers. Magneto-transport measurements in pulsed fields up to 72 Tesla on AHTS spanning the full doping range reveal that the maximum TC occurs at L/d = 0.75 (yielding average hole doping $\langle\delta\rangle = 0.45(1 - L/d) \approx 0.11$), shifted from natural cuprates ($\langle\delta\rangle \approx 0.15$). The IMT occurs at $L/d$ = 0.67 ($\delta \approx 0.15$) in AHTS versus $\langle\delta\rangle \approx 0.18$ in natural cuprates. Both $T_C$ and $\mu_0 H_{C2}$ peak at $L/d \approx 0.75$, while $T_C \xi_0$ peaks at $L/d$ = 2/3, where the Lifshitz transition from donut to cylinder Fermi surface occurs in the second subband. The evolution of $\mu_0 H_{C2}$(T), the Ginzburg-Landau coherence length $\xi_0$, and the product $T_C \xi_0$ from underdoped to overdoped regime reveals a BEC-to-BCS crossover, consistent with the Fano-Feshbach shape resonance in a superlattice of quantum wells predicted by the Bianconi-Perali-Valletta theory for multigap superconductivity with a key role of Rashba spin-orbit coupling.

## 2. Materials

The samples investigated in this work are artificial high-$T_C$ superlattices formed by alternating atomic layers of stoichiometric Mott insulating $La_2CuO_4$ (LCO) and overdoped metallic $La_{1.55}Sr_{0.45}CuO_4$ (LSCO) [42-47]. As schematically illustrated in Fig. 1a, the heterostructures realize a three-dimensional superlattice of superconducting quantum wells, where undoped LCO layers with thickness $L$ are periodically intercalated by metallic LSCO spacer layers with thickness $W$.
The resulting superlattice period is therefore $d=L+W$, while the electronic properties are controlled by the geometrical conformational parameter $L/d$. Superconductivity emerges at the interfaces through charge transfer from the metallic LSCO layers into the nominally undoped LCO blocks.

By varying the $L/d$ ratio, it is possible to tune the average hole concentration and quantum confinement of the space charge at the interface without introducing chemical substitution into the superconducting layers. The formal doping was estimated via the relation $\langle\delta\rangle = 0.45(1 - L/d)$ allowing systematic exploration of the superconducting dome from the underdoped regime ($\delta < 0.15$), which is revealed to be a BEC-like regime in the rising edge of the dome, and the overdoped regime ($\delta > 0.15$) revealed to be a BCS-like regime in the decreasing edge of the dome in agreement with a Fano-Feshbach resonance scenario.
Figure 1b reports the normalized sheet resistance as a function of the temperature for the investigated LSCO/LCO superlattices. A clear evolution of the normal-state transport is observed across the series, ranging from the pseudogap regime with the low-temperature Kondo-like upturns for $\delta < 0.15$ to the strange metal Planckian dissipation regime, with a first extended linear-in-temperature resistivity component dominant at the critical doping $\delta = 0.15$, highlighted by the dashed line. As shown in Fig. 1c, the superconducting critical temperature reaches its maximum value in the range $0.66 < L/d < 0.7$, corresponding to doping 0.10-0.12. The identification of the lineshape of the superconducting dome is due to the key advantage of the artificial heterostructures, where the superconducting LCO quantum wells remain structurally stoichiometric and largely free from the disorder associated with conventional chemical doping like CDW puddles.

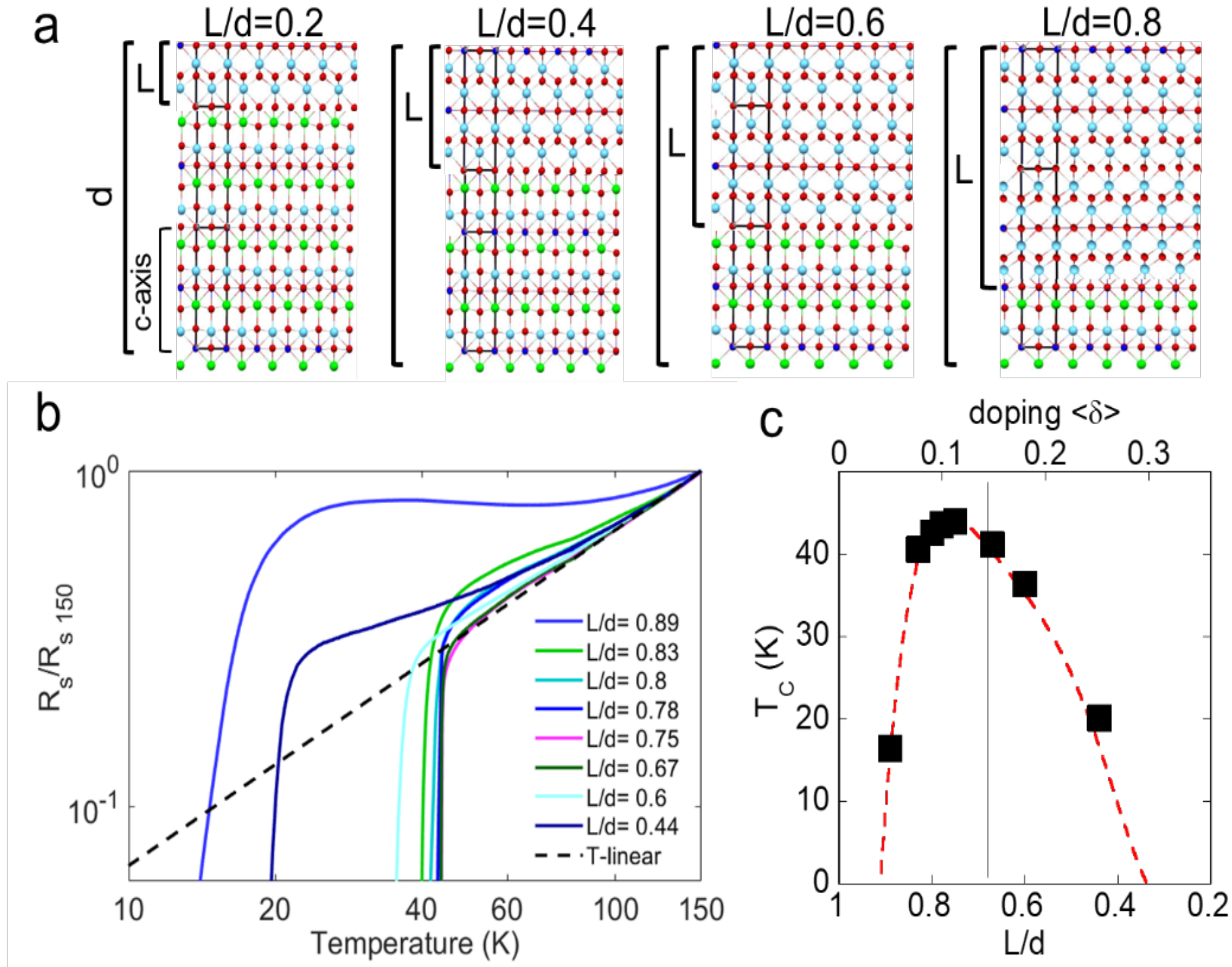

**Figure 1.** (a) Crystal structure of an AHTS superconductor composed of a 3D superlattice of quantum wells made of stoichiometric LCO units of thickness $L$, intercalated by chemically doped normal metal units LSCO of thickness $W$ with period $d = L + W$. Conformational parameter $L/d$ yields the optimum critical temperature in the range $0.6 < L/d < 0.7$, (b) Normalized sheet resistance as a function of temperature for LSCO/LCO superlattices with different $L/d$ ratios measured in this work. The dashed line represents the $T$-linear resistivity expected for the Planckian regime in the range $0 < T < 150$ K. (c) Superconducting dome, $T_C$ as a function of $L/d$, for the AHTS superlattices measured in this work.

## 3. Methods

Magneto-transport measurements were performed at the pulsed high magnetic field facility of Los Alamos National Laboratory using the 72 T Half-Duplex Magnet, which provides non-destructive magnetic field pulses up to 72 T with an 80 ms pulse duration. The magnet is equipped with a 9 mm bore cryogenic sample space, enabling transport measurements down to temperatures as low as 0.5 K. In the present experiments, two samples were mounted simultaneously in the probe, allowing direct comparison under identical magnetic field and thermal conditions.

Electrical resistance was measured using a standard four-probe high-frequency AC technique, with the excitation current applied parallel to the superconducting planes ($ab$-plane transport geometry), while the magnetic field was oriented perpendicularly to the layers ($H \parallel c$-axis). The excitation frequency was carefully optimized around 30 kHz, high enough to minimize low-frequency noise and improve signal-to-noise ratio, while remaining sufficiently low to avoid significant reactive (capacitive and inductive) contributions to the measured voltage signal. This optimization ensured that the detected response remained predominantly resistive over the full magnetic field range, enabling accurate determination of the longitudinal magnetoresistance under rapidly varying field conditions.

The temporal profile of the magnetic field pulse is particularly well suited for high-resolution transport measurements. Following the rapid rise to maximum field, the magnetic pulse enters a significantly slower decay phase, characterized by a reduced sweep rate, smoother field evolution, and the absence of abrupt temporal features associated with the rising edge of the pulse. Owing to these conditions, the decreasing-field branch provides a substantially improved signal-to-noise ratio, reduced inductive pickup, and lower sensitivity to dynamic artifacts, enabling more accurate resistance measurements under quasi-steady high-field conditions. For this reason, all magneto-transport data presented in this work were acquired and analysed using the descending part of the magnetic field pulse. The voltage response was recorded using high-speed synchronized lock-in detection electronics, allowing precise reconstruction of the full $R(\mu_0 H)$ curves at fixed temperatures over the entire field range investigated.

Measurements were performed over a broad temperature range spanning the superconducting transition and extending down to the base temperature of 1.5 K. This enabled the systematic determination of both isothermal magnetoresistance curves $R(\mu_0 H)$ and field-suppressed normal-state resistance curves $R(T)$. The upper critical field, $\mu_0 H_{C2}(T)$, was extracted from the resistive transitions using multiple criteria, corresponding to the 50% and 100% of the normal-state resistance, thus providing a robust estimate of the superconducting phase boundary across the entire doping range investigated.

## 4. Results and discussion

We begin by discussing the magneto-transport measurements performed on a series of eight samples spanning the doping range across the superconducting dome, with particular focus on underdoped and near-optimal. doping compositions. The field dependence of the resistance, $R(\mu_0 H)$, measured at different temperatures is shown in **Fig. 2**. A striking feature emerges upon comparing samples across the doping series as a function of the superlattice parameter $L/d$. For underdoped compositions ($\delta < 0.15$), the $R(\mu_0 H)$ curves measured at different temperatures systematically cross at a well-defined field, forming an isosbestic-like point. The position of this crossing shifts to lower magnetic fields upon reducing doping, i.e., moving deeper into the underdoped side of the dome, and progressively moves toward higher fields as doping approaches the optimal values.

Remarkably, for samples with δ > 0.15, the crossing point is no longer observed within the investigated field and temperature range, and the $R(\mu_0 H)$ curves do not intersect. This indicates that in AHTS superlattices the critical point for the transition from the Kondo-like dominated regime to the strange metal regime is δ > 0.15 while in natural cuprates it is known to be δ ≈ 0.18 as measured by using high-field magnetoresistance [38,39]. This qualitative change indicates a profound modification of the normal-state transport properties in cuprate AHTS with respect to natural HTS. The existence of the isosbestic point is related with the presence of a low-temperature Kondo-like upturn in the normal-state resistance once superconductivity is suppressed by the magnetic field; conversely, its disappearance is associated with a monotonic metallic resistivity characteristic of a *strange-metal* regime with a Planckian linear-in-$T$ component coexisting with a Fermi liquid component.

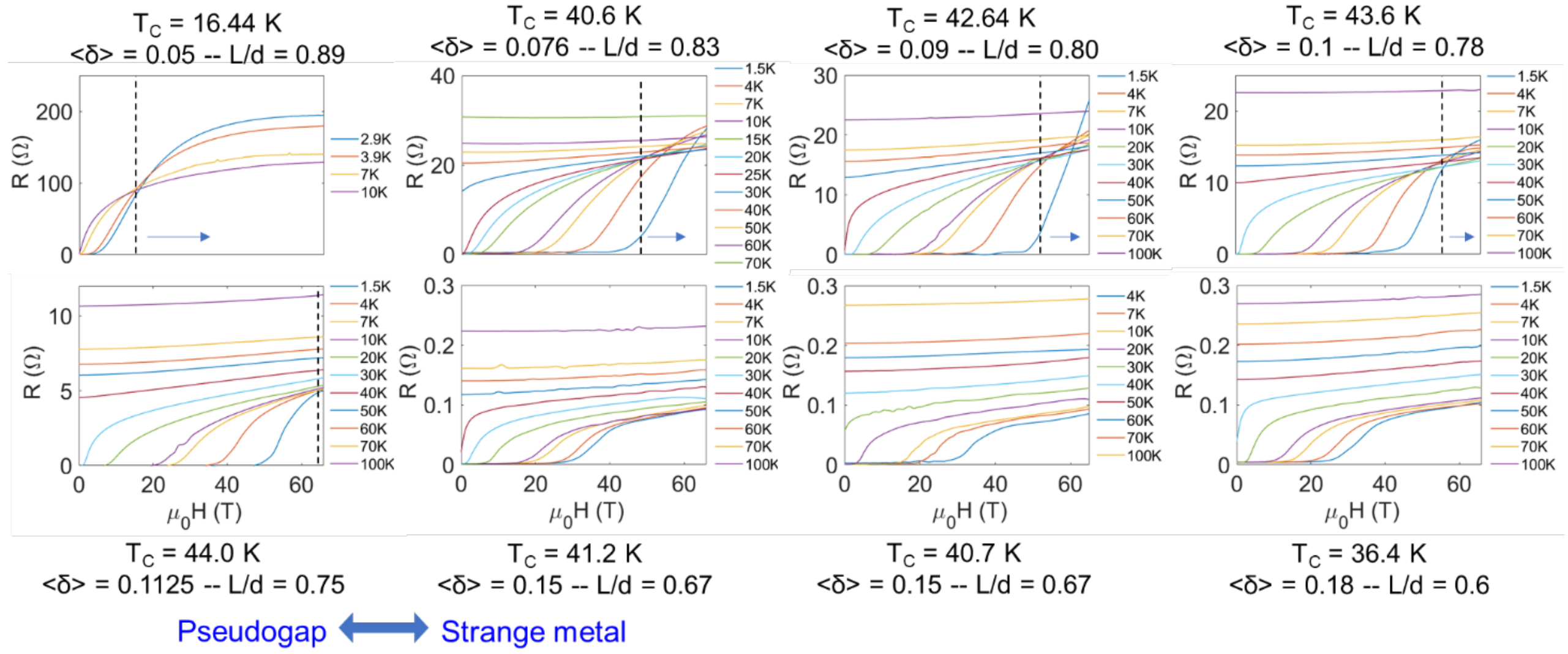

**Figure 2.** Resistance $R(\Omega)$ as a function of the temperature in eight samples with different $T_C$ and $L/d$ (<δ>) values (indicated at the top of the upper panels), measured under magnetic field $\mu_0 H$ up to 72 T at the NHMFL-LANL facility.

To expose this crossover, we extracted the temperature dependence of the resistance, $R(T)$, collected at fixed magnetic fields by taking vertical cuts of the $R(H)$ data, see **Fig. 3**. These curves provide a direct view of the evolution of the normal-state transport as a function of both field and doping. For samples exhibiting the isosbestic point in $R(H)$, the corresponding $R(H)$ curves display a clear upturn at low temperatures once superconductivity is suppressed by the magnetic field.
This low-temperature increase of the resistivity is progressively reduced upon increasing doping and eventually disappears in the sample at the critical doping $p \sim 0.15$, where $R(T)$ remains linear down to the lowest accessible temperatures. The low-temperature upturn observed here agrees with a logarithmic increase expected from Kondo-like scattering mechanisms, as previously reported in our earlier work on the normal resistance at $T > T_C$ [44] given by

$$\frac{R(T)}{R(T=150K)} = \frac{T}{150} + r_0 + AT^2 + BT^5 + \frac{R_{0K}}{\left\{1+\left(2^{\frac{1}{s}}-1\right)\left(\frac{T}{T_K}\right)^2\right\}^s} \tag{1}$$

where $T_K$ is a characteristic Kondo temperature In **Figure 4a** we show the normalized resistances measured as a function of $T$ under high $\mu_0 H = 60$ T for the eight samples studied here. The $T_k$ and $R_{0k}+r_0$ parameters, obtained by modelling these resistances via Eq.1, are shown in **Fig. 4b**. It is worth mentioning that, even when $T_K > T_C$, the resistance minimum is uncovered only under a sufficiently large magnetic field, when superconductivity has been suppressed.

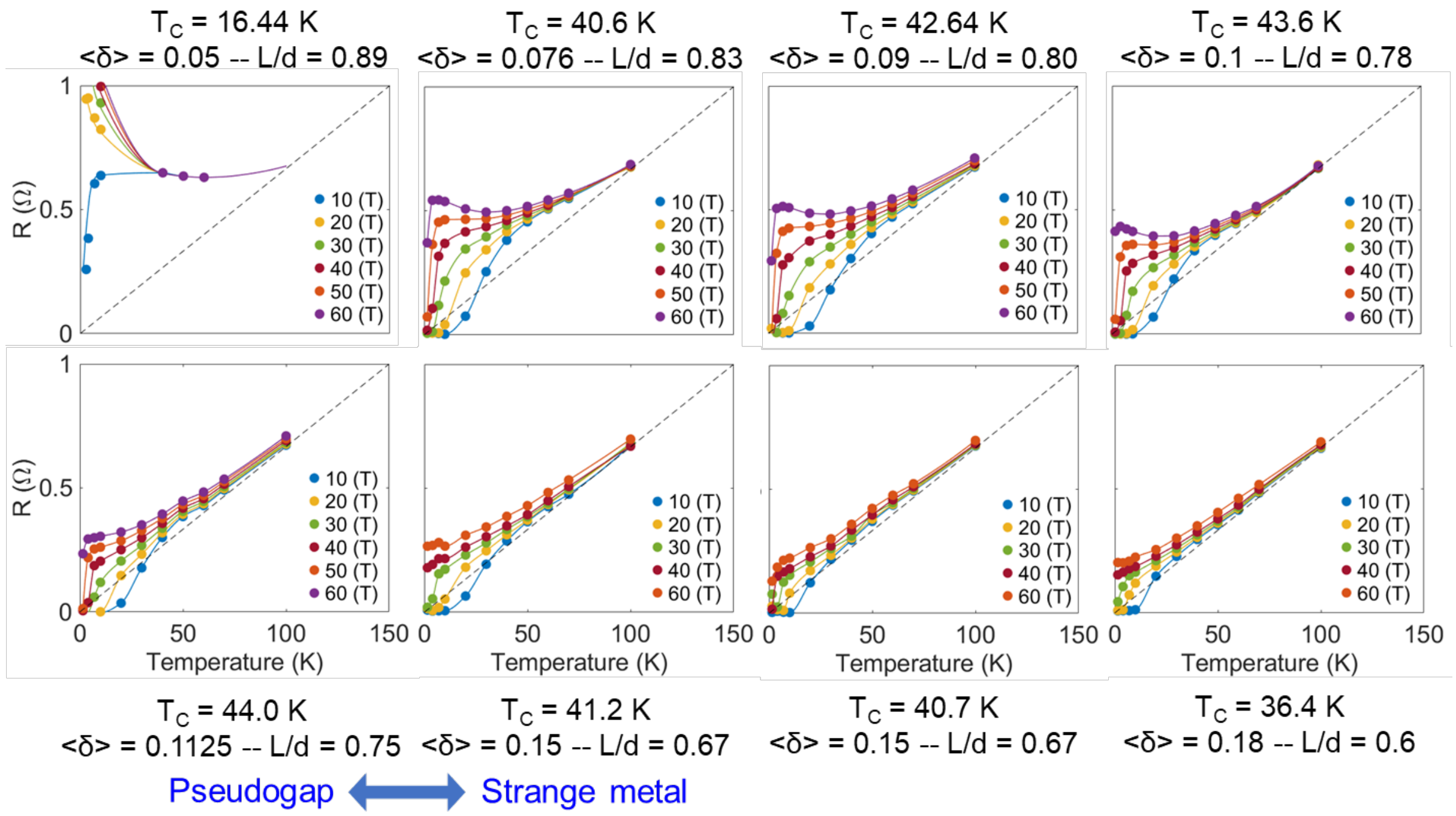

**Figure 3.** Resistance as a function of temperature under the indicated magnetic fields for the eight samples studied here. Values of d are indicated at the top of the upper panels and bottom of lower panels. We observe the low-temperature upturn in resistance characteristic of the pseudogap phase with a Kondo-like component coexisting with Planckian dissipation, dominant at δ=0.15

The systematic disappearance of both the low-temperature upturn in *R(T)* and the crossing point in *R(H)* upon increasing doping indicates that this additional scattering channel weakens progressively. Near optimal doping, the transport is instead dominated by a robust linear-in-temperature resistivity, characteristic of a strange-metal regime.

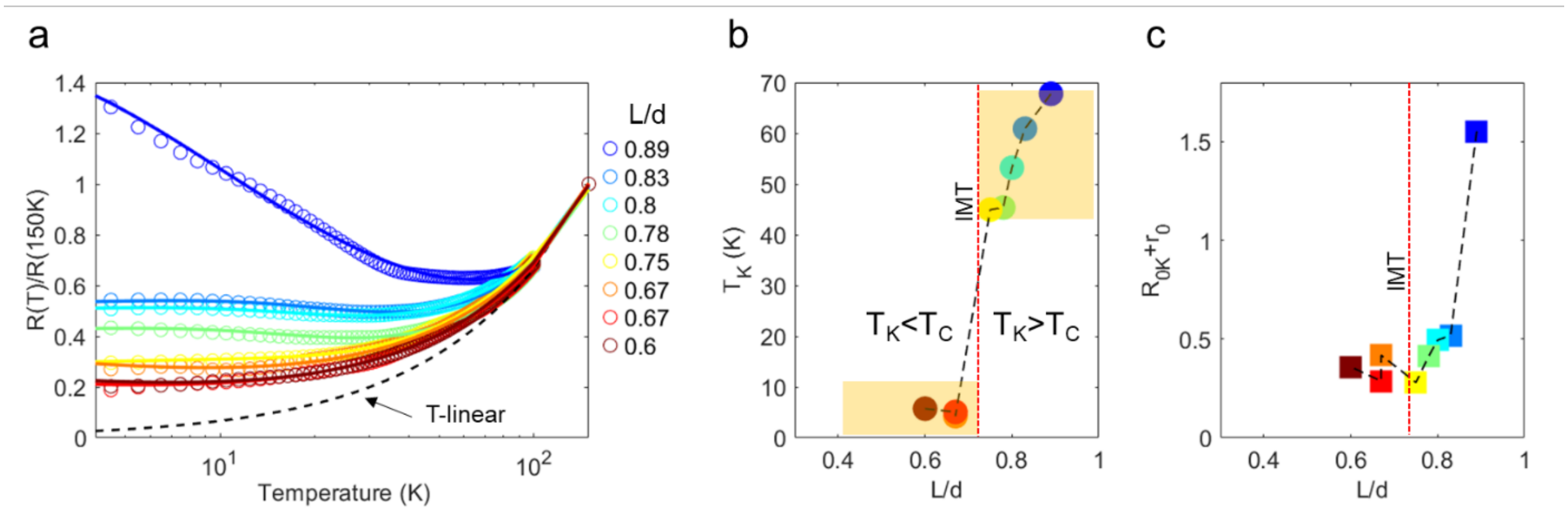

**Figure 4** (a) Normal state resistance (open circles) normalized by the value of the resistance at T = 150 K or *R*(150K) as a function of *T* under $\mu_0 H$ = 60 T for the eight samples studied here, alongside the modelled curves (continuous lines) via Eq. 1. Different values of *L*/*d* are indicated by different colors. (b) $T_k$ and $R_{0k}$+$r_0$ parameters extracted through the fitting procedure. We can distinguish the strong Kondo-like regime, where $T_K > T_C$, and the weak Kondo-like regime with $T_K < T_C$.

Upon further increasing doping, a crossover to strange metallic state is observed, where a Fermi-liquid-like behaviour emerges at low temperatures, characterized by both a quadratic temperature dependence of the resistivity, $\rho(T) \sim T^2$, due to coherent quasiparticle scattering coexisting with Planckian low-temperature *T*-linear resistivity regime. This evolution is consistent with the phenomenology observed in cuprates near optimal doping, where Planckian dissipation leads to an extended $T$-linear resistivity regime, followed at higher doping by the gradual increase of a Fermi-liquid-like component coexisting with the Planckian component.

To gain deeper insight into the evolution of the normal and superconducting states across the doping levels, we extracted the upper critical fields $\mu_0 H_{c2}$ using two criteria, corresponding to 50% and 100% of the resistive transition $\mu_0 H_{c2}^{50}$ and $\mu_0 H_{c2}^{100}$ (also called $H_{50}$ and $H_{100}$), as summarized in **Figs. 5** and **6**. The color-contour maps of the derivative $dR/d(\mu_0 H)$ as a function of $\mu_0 H$ and $T$ (Fig. 5) provide a comprehensive view of the resistive transition landscape for each sample. The black dots and thick lines represent $\mu_0 H_{c2}^{100}$, dashed lines indicate the irreversible field $\mu_0 H_{cirr}$, and open circles correspond to the maximum of the derivative modeled by a Fano line shape, yielding $\mu_0 H_{c2}^{50}$, so a clear and systematic doping dependence emerges.

For all the δ values investigated, both $\mu_0 H_{c2}^{50}(T)$ and $\mu_0 H_{c2}^{100}(T)$ exhibit a pronounced upward curvature upon lowering temperature, as shown in Fig. 5(a) and (b). This behavior constitutes strong deviations from the single-band Werthamer–Helfand–Hohenberg (WHH), which predicts a downward or at most linear curvature near $T_C$. Instead, the observed upward curvature is a hallmark of multiband superconductivity, where the interplay between bands with different gap magnitudes and diffusivities produces a characteristic positive upward curvature for the phase boundary [46,47]. This confirms the persistence of multigap physics across the entire dome, unlike our previous study [47] limited to four samples at the dome edges. Taken together, these results allow us to identify distinct regimes across the phase diagram, namely: a) *pseudogap regime characterized by a Kondo like low temperature upturn, b) Planckian*, and c) *strange-metal* regimes, each characterized by a specific interplay between normal-state transport and superconducting properties.

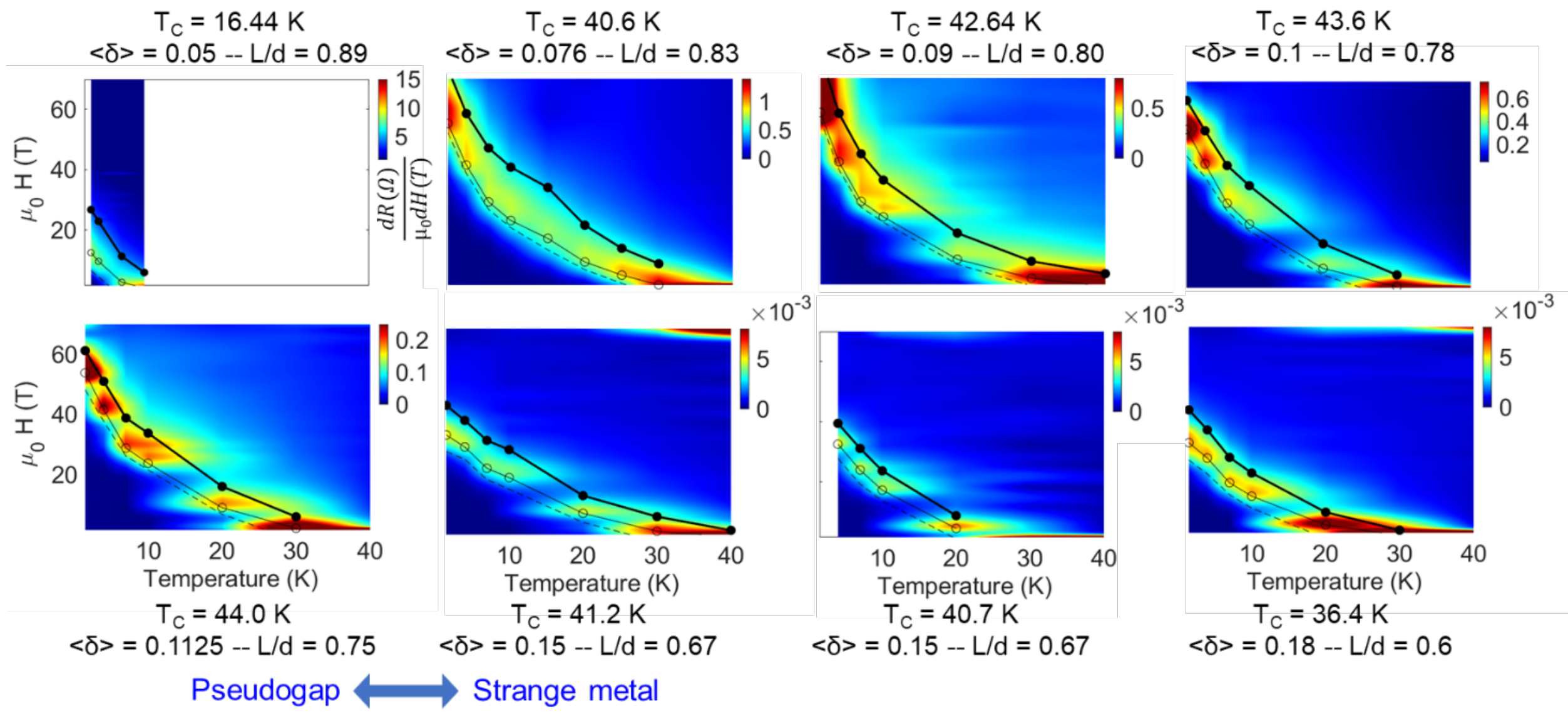

**Figure 5**. Color contour plots of $dR/\mu_0 d(H)$ as a function of $\mu_0 H$ and $T$, for the samples with different L/d ratios. Black dots alongside the thick black line represent the upper critical magnetic fields, namely, $\mu_0 H_{100}$. The dashed black lines are the irreversible magnetic fields $\mu_0 H_{Cirr}$. Open circles represent the maximum of the derivative modeled by the Fano line shape.

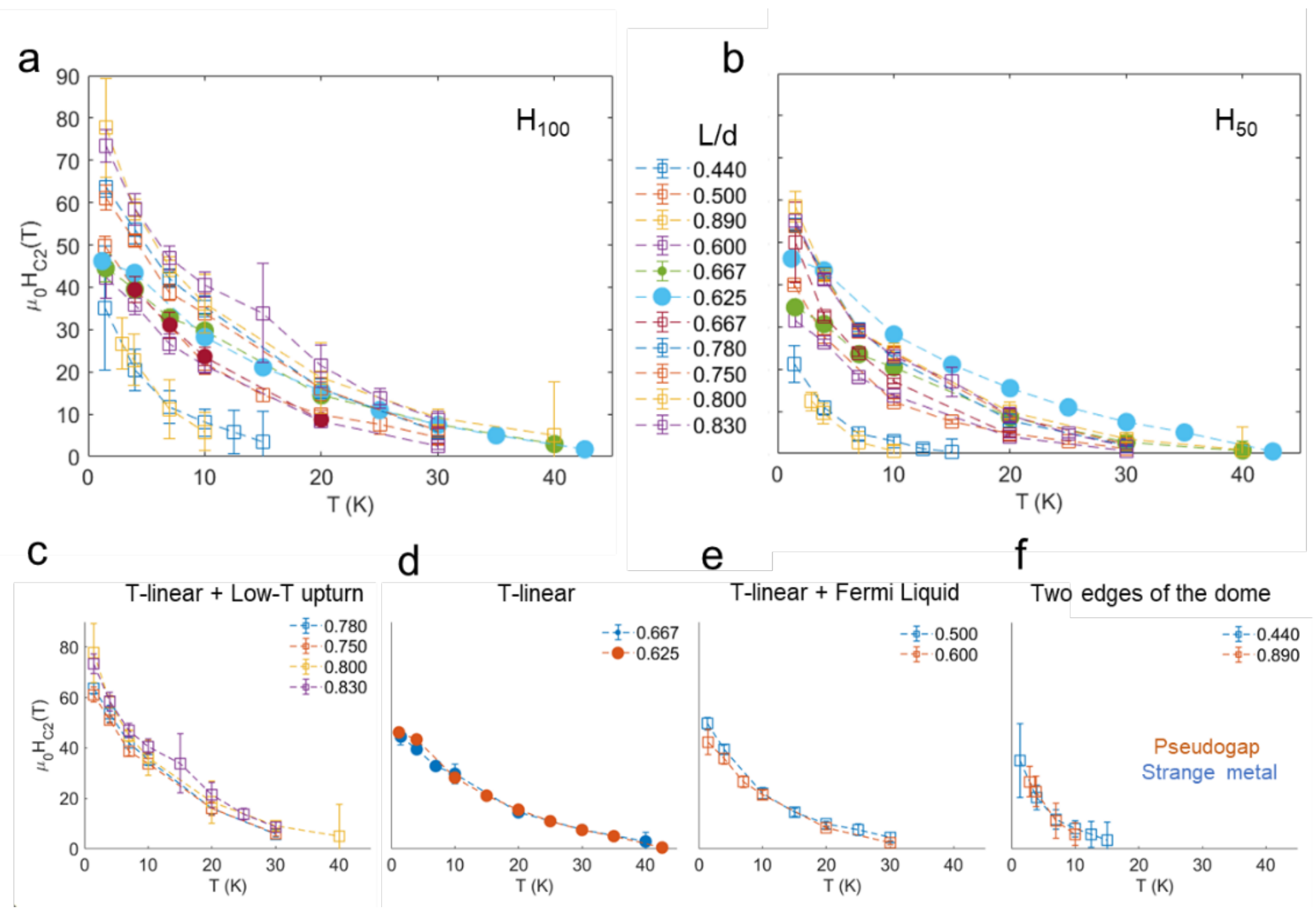

**Figure 6.** Critical magnetic field $\mu_0 H_{c2}$(T) as a function of $T$ for the eight samples studied here. We report both the $\mu_0 H_{100}$ (a panel) and $\mu_0 H_{50}$ (b panel) For completeness, we report also the $\mu_0 H_{c2}(T)$ values (full circles) obtained for the three samples having $L/d$ = 0.625, 0.67, in the optimum doping regime and data from Campi *et al.* [5]. (c) Data points corresponding to different $\mu_0 H_{c2}(T = 0)$ for all four regimes. We observe changes in the $\mu_0 H_{c2}(T)$ values as the system crosses the distinct regimes.

The grouping of the $\mu_0 H_{c2}$ curves into these regimes, see Fig. 6 (c)-(f), reinforces the view that the crossover observed in the normal state is intimately connected to the underlying superconducting resonating pairing mechanism predicted by the Bianconi-Perali-Valletta (BPV) theory in quasi-2D interfacial electron gas with Rashba spin-orbit at the Lifshitz electronic transition from a donut shaped Fermi surface to a corrugated cylindrical Fermi surface appearing near the band edge of the second subband.

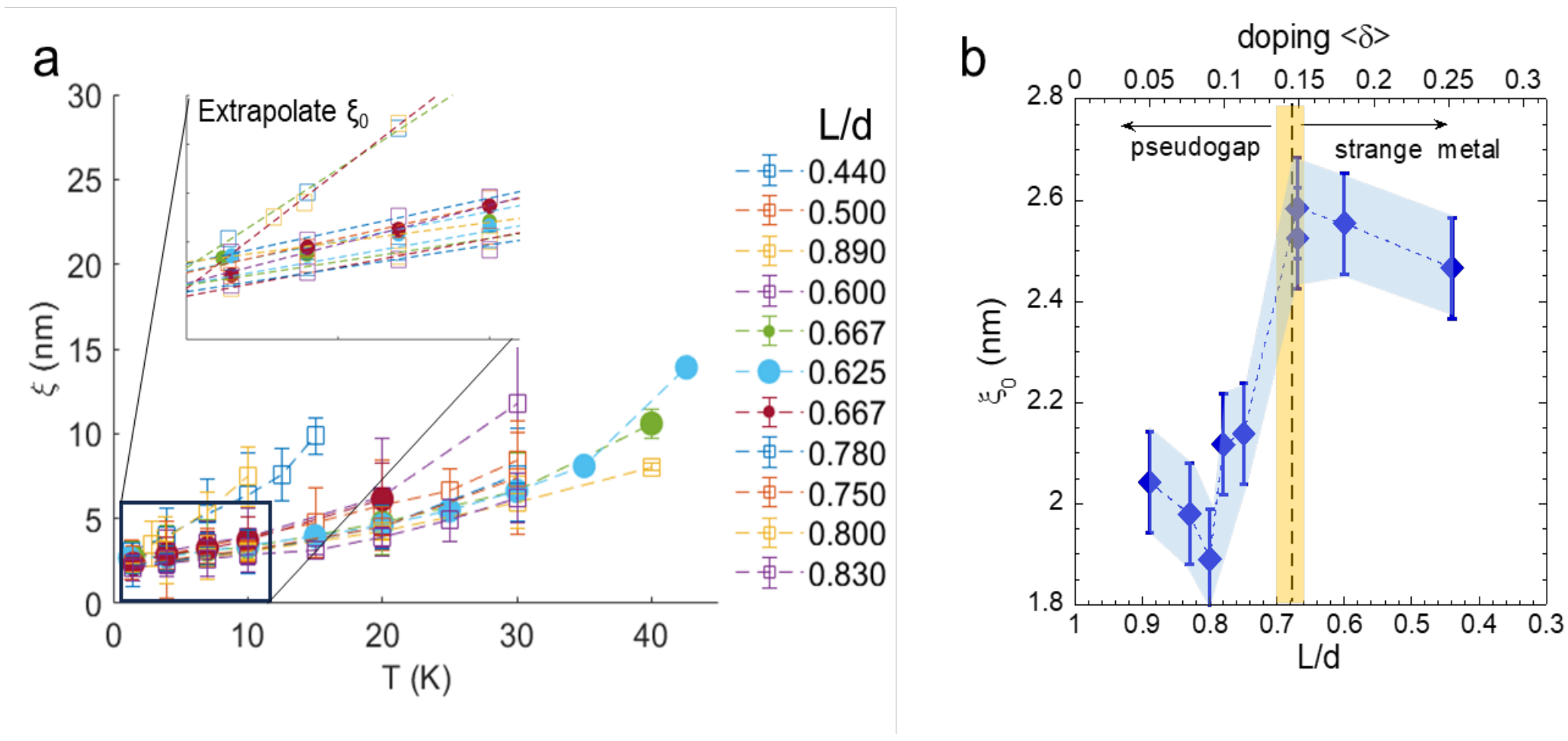

**Figure 7.** (a) Temperature dependence of the coherence length x. Zoom in of the $\xi$-$T$ plane used to extrapolate the trend toward zero temperature to obtain the intrinsic $\xi_0$. (b) $\xi_0$ as a function of both the $L/d$ parameter and the average doping δ =0.45 $(1 - L/d)$. The $\xi_0$ values have been extrapolated from the $\xi(T \rightarrow 0$ K) curves highlighted by the black rectangle in panel (a).

The upper critical field in the zero-temperature limit is directly related to the superconducting Ginzburg-Landau coherence length through the relation.

$$\xi(T)^2 = \frac{\Phi_0}{2\pi H_{c2}(T)} \quad (2)$$

where $\Phi_0$ is the flux quantum. The temperature-dependent Ginzburg-Landau coherence length $\xi(T)$, extracted from the measured $\mu_0 H_{c2}^{100}(T)$, is reported in Fig. 7(a) for representative samples. The extrapolation of $\xi(T)$ to zero temperature yields the intrinsic $\xi_0$, plotted in Fig. 6(b) as a function of both $L/d$ and the average doping $\delta$. A clear trend emerges: $\xi_0$ decreases monotonically upon reducing doping, reaching its minimum value in the underdoped side of the dome. This indicates that Ginzburg-Landau coherence length decreases as the carrier density is reduced, a behavior qualitatively consistent with the system moving toward the BEC side of the BCS-BEC crossover.
The coherence length controls the robustness of the superconducting phase against fluctuations of the superconducting order parameter. The strength of the fluctuations is quantified by the Ginzburg parameter $G_i$ that, in two dimensions, is inversely proportional to the coherence length:

$$G_i = 1/(k_F \xi_0) \quad (3)$$

A larger $\xi_0$ favors higher critical temperatures provided that the underlying electronic dispersion remains unchanged, and pairing is strong enough to boost the pairing temperature $T_C$. The stabilization of higher-temperature superconductivity therefore requires two mechanisms acting at the same time; a strong pairing to enhance the mean-field critical temperature where Cooper pairs start forming, and large coherence lengths (small Ginzburg parameter) to screen the detrimental superconducting fluctuations [56,57,58]. This suppression of fluctuations is particularly important when the system is close to a van Hove singularity. This optimal configuration is extremely difficult to achieve in single-band, single-gap superconducting systems, but it is a natural outcome in two-band, two-gap superconductors, in which one partial condensate is in the BCS regime (large coherence lengths) and the second partial condensate is in the BCS-BEC crossover regime (strong pairing).

The combined analysis of $T_C$, $\mu_0 H_{c2}(0)$, $\xi_0$, and $T_c\xi_0$, summarized in Fig. 8, provides the central result of this work. **Fig. 8(a)** and **Fig. 8b** displays $T_C$ and $\mu_0 H_{c2}(0)$ as functions of both $L/d$ and average δ. The yellow stripe centered at doping $<d> \sim 0.15$ marks the doping value separating the pseudogap regime, where resistivity displays the low-temperature Kondo-like upturn, from the strange metal phase. The theoretical superconducting dome predicted by the BPV calculations [55] are represented as overlayed colormaps for LCO/LSCO superlattices with two different theoretical superlattice periods d = 3nm (panel a) and period d=3.96 nm (panel b) showing their control of the width of theoretical superlattice dome [55].
The superconducting dome of high-$T_C$ natural cuprate crystals is shown in **Fig. 8(c)**; here underdoped side of the superconducting dome is strongly affected by competing electronic orders, in particular charge-density-wave order. CDW correlations, which compete with superconductivity because they partially gap the Fermi surface, reduce the density of itinerant carriers available for Cooper pairing, reconstruct the low-energy electronic structure, and weaken the superfluid stiffness. This competition naturally leads to a reduction of $T_C$ in the underdoped regime. Indeed, in underdoped cuprate crystals it has been shown that the suppression of the CDW order by strain enhances superconductivity, demonstrating the antagonistic relation between the two phases [32-36], and resorts the strange metal behavior.
On the contrary, in the artificial high-$T_C$ superlattices studied here [37-42], the situation appears markedly different, (see Fig. 8c), because these samples are structurally cleaner while the carrier density is controlled by the geometrical parameter $L/d$, instead of chemical substitutional disorder. The underdoped side of the dome is not dominated by the CDW instability that suppresses superconductivity in bulk cuprate [33-36] crystals. As a result, the intrinsic superconducting state can be accessed directly. This is reflected in Fig. 8a and Fig. 8b where both $T_C$ and $\mu_0 H_{c2}(0)$ are strongly enhanced on the underdoped side, instead of being suppressed as commonly observed in bulk HTS crystals. In particular, the large values of $\mu_0 H_{c2}(0)$ imply a short Ginzburg- Landau coherence length with a robust superconducting condensate. Therefore, the underdoped region in these artificial superlattices should not be viewed simply as a weakened superconducting phase, but rather as a regime where strong pairing is favored in the BEC side of the superconducting dome predicted by the shape resonance effect according to the BPV theory.

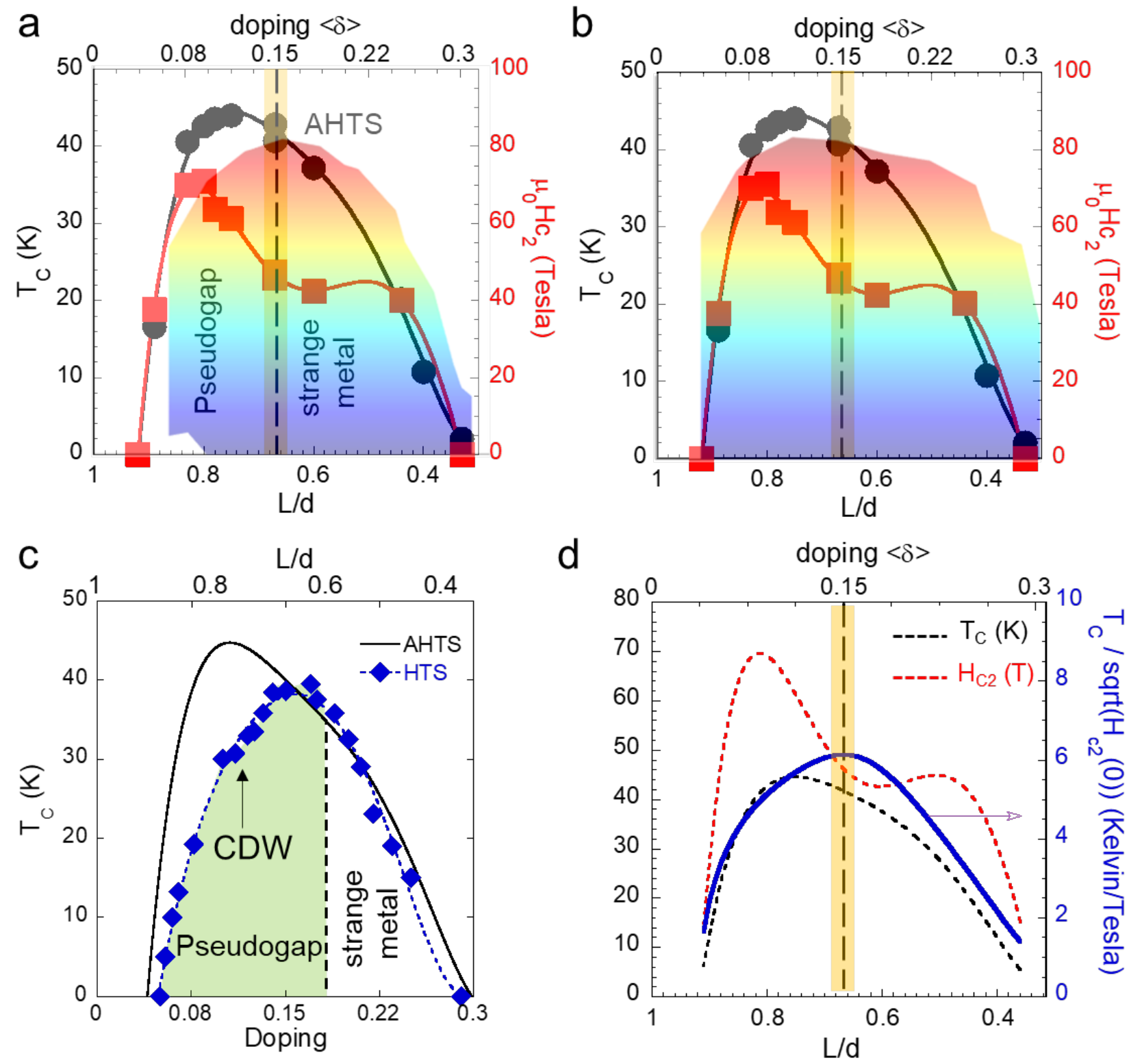

**Figure 8 (a, b)** Upper critical magnetic field, $\mu_0 H_{c2}$, (red squares) as a function of both the $L/d$ parameter and the average doping $\delta = 0.45(1 - L/d)$ compared with the critical temperature $T_C$ (black dots) in the superconducting dome of AHTS. The first stronger peak in $\mu_0 H_{c2}$ occurs for $L/d < 2/3$ in the pseudogap regime, separated by a weaker broad peak in the strange metal regime for $L/d > 2/3$. In AHTSs the strange metal phase occurs for δ>0.15. The yellow stripe at L/d = 2/3 (<δ> = 0.15) separates the multigap BEC-like regime from the BCS-like multigap regime as a function of L/d (<δ>) [55] The overlay colormaps represent the theoretical predictions of the BPV shape resonance theory [55] for LCO/LSCO superlattices with period d = 3nm (panel a) and period d=3.96 nm (panel b). **(c)** Experimental $T_C$ in AHTSs (continuous black line) compared with the experimental superconducting dome of $La_{1-x}Sr_xCuO_4$ (HTC) whose competing CDW phase suppresses the critical temperature with the strange metal phase occurring only for δ > 0.18. **(d)** Ratio $T_C/\sqrt{\mu_0 H_{C2}(0)}$ (blue solid line), which is proportional to the product $T_C \xi_0$, as a function of $L/d$ showing a maximum at the magic ratio $L/d = 2/3$. This value corresponds to the Lifshitz transition from a donut-like to a corrugated cylindrical Fermi surface associated with the second subband as predicted by the BPV theory including spin-orbit coupling as shown in Fig.9 [48-55].

Fig. 8(c) compares the experimental superconducting dome of HTS with the theoretical prediction by the BPV theory [55]. The BPV theory framework successfully captures the multigap character of the superconductivity in these heterostructures and the shape of their superconducting dome. The formulation of Lifshitz transition-driven shape resonances in the BEC side at the rising edge of the superconducting dome does not explicitly account for strong electronic correlations, which become increasingly relevant in the low-density underdoped regime. Correlation-driven physics may explain the deviations observed between the experimental superconducting dome and the BPV prediction. The overall width of the superconducting dome predicted by this theory increases increasing the period d from 3nm (panel a) to the period d=3.96 nm (panel b) of Fig.8, indicating that the global superconducting phase is still governed by quantum confinement and multigap resonance effects, the underdoped side exhibits a clear strengthening of superconductivity, as evidenced by an enhanced $T_C$, larger $\mu_0 H_{c2}(0)$, and a shorter Ginzburg-Landau coherence length. This low-density experimental enhancement is captured by increasing the superlattice period and the formulation of the BPV model with longer superlattice period and suggests that, in the clean limit realized by these artificial heterostructures, strong correlations can reinforce superconductivity on the underdoped side, revealing an intrinsic BEC-BCS crossover-like regime.

Finally, **Fig. 8d** shows the ratio $Tc/\sqrt{\mu_0 H_{c2}(0)} \propto Tc\xi_0$ which combines the energy and length scales of the superconducting transition, set by $T_C$ and $\xi_0$, yielding a microscopic measure of the coherence and of the pairing, averaging their effect to quantify the stability of the superconducting state. Remarkably, this ratio exhibits a maximum at $L/d = 2/3$, which corresponds to the Lifshitz electronic topological transition in the second appearing subband from a donut shape shown in Fig. 9 panel (1), to resulting emergence of a new second corrugated cylindrical Fermi surface **Fig. 9** panel (3) where the BPV theory calculations predict the enhanced Tc due to "shape resonance" in the pair transfer exchange term from the first to the second subband [55].

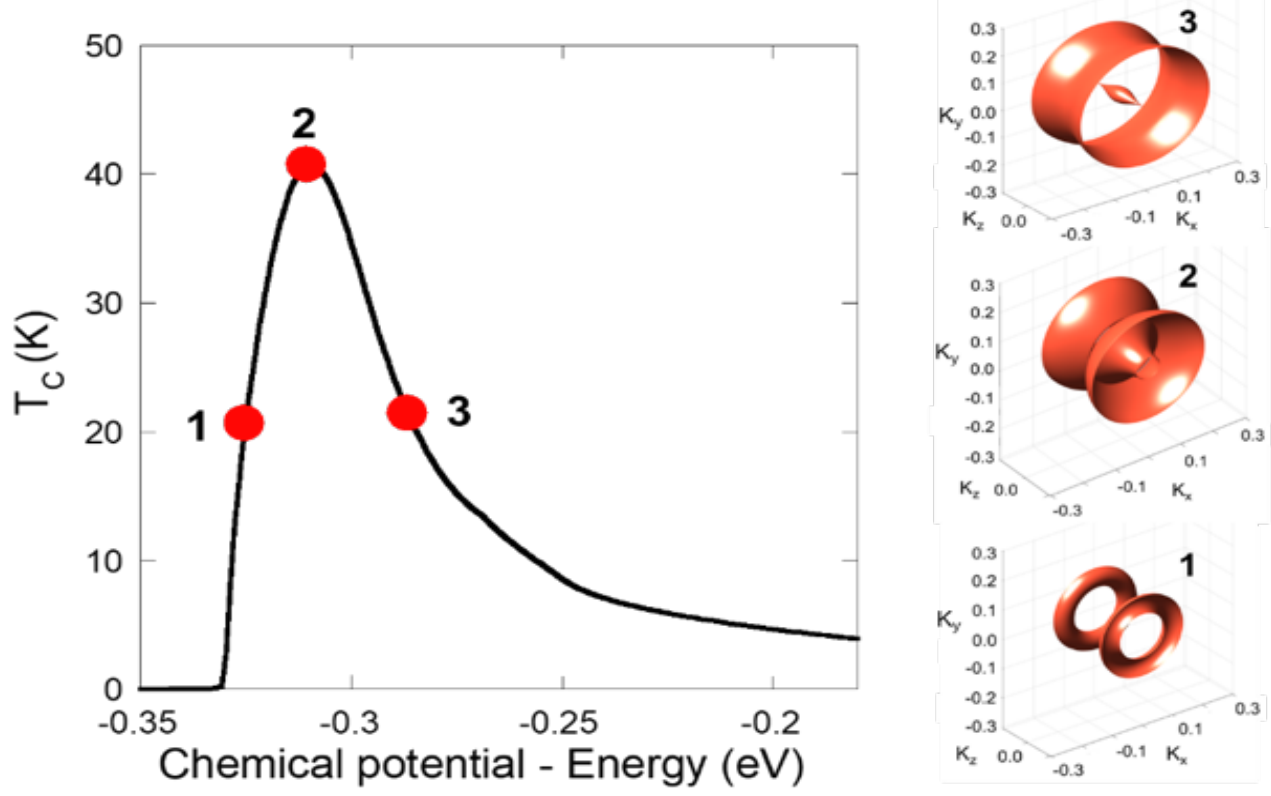

**Figure 9** Theoretical first principles superconducting dome for an artificial high $T_C$ superlattice (AHTS) of quantum wells at fixed magic geometry ratio L/d=2/3 calculated by tuning of the chemical potential near the topological Lifshitz transition from a donut-like Fermi surface to a corrugated cylindrical Fermi surface in the second subband. At the rising edge of the dome (1) the multigap superconducting phase is in the BEC-like pairing regime, while in the drop edge (3) it is in the multigap BCS-like regime [55].

## 4. Summary and conclusions

Superconductivity emerging from doped Mott insulators has long remained as one of the most challenging problems in condensed matter physics, largely because the intrinsic electronic phases are usually entangled with the disorder introduced by the chemical doping. In conventional cuprates, the same dopants that provide mobile carriers also generate impurity scattering, local lattice distortions, nanoscale electronic inhomogeneity, and pinning of competing orders such as charge-density waves. As a result, the underdoped region of the phase diagram is particularly difficult to interpret: the suppression of $T_C$, the reduction of $\mu_0 H_{C2}$, and the emergence of pseudogap behavior may reflect not only the intrinsic physics but also disorder, induced localization, and competing phases. The present artificial LSCO/LCO superlattices offer a fundamentally different route. Here, the carrier density is controlled geometrically through the superlattice architecture, while the superconducting LCO quantum wells remain chemically clean. This makes it possible to access a much less perturbed version of the normal and superconducting state in the underdoped side of the dome.

In this cleaner system, our results reveal that the low-density regime is not simply a weak or degraded superconducting state. On the contrary, the enhancement of $\mu_0 H_{C2}(0)$, the shortening of the coherence length, and the persistence of robust superconductivity indicate the emergence of a strongly correlated electronic phase where pairing in the BEC-BCS crossover regime remains remarkably effective in the rising edge of the superconducting dome.

The comprehensive magneto-transport study across eight samples spanning the full doping range ($\delta \approx 0.08$–$0.18$), using pulsed magnetic fields up to 72 T, provides systematic evidence for:

- Persistent multiband superconductivity across the entire dome, demonstrated by the upward curvature of *$\mu_0 H_{C2}(T)$* for all *L/d* ratios, in strong deviation from single-band WHH predictions.
- Distinct components transport regimes:
  - *pseudogap phase for $\delta$ < 0.15* made of two components: a first, characterized by a first Kondo-like component ($T_K$>$T_C$ showing low temperature resistance upturn) and a second, T-linear Planckian component
  - *dominant Planckian* component, for $\delta \approx 0.15$ with T-linear resistance regime with a minor Kondo-like component with $T_K < T_C$
  - *strange metal* phase $\delta \approx 0.15$ made of two components: first, a Fermi-liquid and second, a Planckian component with the crossover occurring at in AHTS rather than $\delta \approx 0.18$ in natural cuprates.
- Quantum geometry optimization, where the ratio $T_C/\sqrt{\mu_0 H_{C2}(0)} \propto T_C \xi_0$ peaks at the theoretical magic value *L/d* = 2/3, while *$\mu_0 H_{C2}(0)$* maximum appears at *L/d* ≈ 0.75.

We find that the maximum $T_C$ occurs at *L/d* ≈ 0.75 ($\delta \approx 0.11$), shifted relative to natural bulk cuprates ($\delta \approx 0.15$), while the low-temperature IMT defined by the disappearance of the Kondo-like resistivity upturn and the onset of purely linear-in-T behavior in R(T), occurs at *L/d* ≈ 0.67 ($\langle\delta\rangle \approx 0.15$), compared to $\delta \approx 0.18$ in natural cuprates. The systematic evolution of *$\mu_0 H_{C2}(T)$*, $\xi_0$, and $T_C\xi_0$ across underdoped, optimally doped, and overdoped regimes reveals a crossover from a BEC-like pairing regime on the rising edge of the dome to a BCS-like regime on the drop edge, consistent with the Fano-Feshbach shape resonance.

This work provides experimental validation for the BPV theory, which proposes that the optimization of high-temperature superconductivity can be achieved by controlling the multi-gap pairing mechanism involving two coupled condensates near a topological Lifshitz transition. This scenario involves a Fano – Feshbach "shape resonance" between two different superconducting condensates coexisting in the same quantum complex system: a weak-coupling BCS condensate and a strong-coupling BEC-BCS crossover condensate, which resonate together maximizing and stabilizing $T_c$ According with the BPV theory first proposed in 1996, the superconducting dome is generated by the Fano-Feshbach shape resonance for pair transfer between two subbands [48-55] which is dominated by BEC-like behavior in the *underdoped regime* on the *rising edge* of the superconducting dome and a BCS-like behavior in the *overdoped regime* on the *drop edge* side of the dome where $T_C$ and the superfluid density decrease. This agrees with the experimental observation of the underdoped phase in the BCS-BEC crossover observed by Uemura in a large set of unconventional high-$T_c$ superconductors [59-60] and low temperature T-linear resistance at optimum doping [61,62].

The present new experimental results on the emergence of three phase in transport properties in AHTS a) Kondo-like dissipation with low temperature resistance upturn on the rising edge of the dome, b) dominant Planckian regime at the top of the dome and c) strange metal regime on the drop edge of the dome, which can be compared with experiments on their emergence in natural cuprate perovskites [63-66].

Finally, we remark that our work provides compelling experimental results shedding light on the underdoped *pseudogap phase* of superconducting cuprates formed by a *first component in the Planckian reg*ime coexisting with a *second component with Kondo-like dissipation* in agreement with Ayres experimental results on two components in the strange metal phase [66,67] and supporting the theory of Chung, C. H. on the mechanism for quantum-critical Planckian metal phase and Kondo-like dissipation in high-temperature cuprate superconductors [68].

In conclusion, we have shown that artificial high-Tc superlattices of quantum wells grown by molecular beam epitaxy (MBE) [40-43] show a superconducting dome of a clean doped Mott insulator by tuning the electronic states near the Fermi energy by the geometry ratio L/d in agreement with predictions of BPV theory [55] and providing new info on the long-standing experimental investigations on unconventional superconductivity in strongly correlated doped Mott insulator remaining elusive since 1986.

**Acknowledgements:**

The authors acknowledge the National High Magnetic Field Laboratory supported by the National Science Foundation through NSF/DMR-2128556, the State of Florida, and the MagLab's Pulsed Field Facility, US Department of Energy. A.B. and G.C. acknowledge the Superstripes - APS association and the CNR project DCM.AD006.562 "Functional Disorder in Materials and Complex Systems" for supporting this work. L.B. acknowledges support from the US DoE, BES program through award DE-SC0002613 US. S.C. acknowledges support from the University of Rome Sapienza, under the projects Ateneo 2023 (RM123188E830D258), Ateneo 2024 (RM124190C54BE48D), and Ateneo 2025 (RP125199B9FDBFE4).